\documentclass[12pt,a4paper]{article}
\usepackage{a4wide}
\usepackage{latexsym}
\usepackage{epsf}
\usepackage{amssymb,amsfonts,amsmath}
\numberwithin{equation}{section}
\def\be{\begin{equation}}
\def\ee{\end{equation}}
\def\bea{\begin{eqnarray}}
\def\eea{\end{eqnarray}}
\def\pd{\partial}
\def\a{\alpha}
\def\b{\beta}
\def\g{\gamma}
\def\d{\delta}
\def\m{\mu}
\def\n{\nu}

\def\l{\lambda}

\def\r{\rho}
\def\s{\sigma}
\def\e{\epsilon}

\def\bg{\bar{g}}
\def\bn{\bar{\nabla}}
\def\br{\bar{R}}
\def\bp{\bar{\phi}}
\def\tr{{\rm tr\,}}

\begin{document}

\begin{titlepage}

\begin{flushright}
IFT-UAM/CSIC-08-31\\[6em]
\end{flushright}

\begin{center}
{\Large\bfseries Ultraviolet behavior of transverse gravity}\\[2em]
{\sc Enrique \'Alvarez, Ant\' on F. Faedo and J.J. L\'opez-Villarejo}\\[1em]
IFTE UAM/CSIC, M\'odulo C-XVI, Universidad Aut\'onoma\\
Cantoblanco, 28049 Madrid, Spain\\[5em]
\end{center}

\begin{abstract}
The structure of the divergences for transverse theories of gravity is studied to one-loop order. These theories are invariant only under those diffeomorphisms that enjoy unit Jacobian determinant (TDiff), so that the determinant of the metric transforms as a true scalar instead of a density. Generically, the models include an additional scalar degree of freedom contained in the metric besides the usual spin two component.  When the cosmological constant is fine tuned to zero, there are only two theories which are on shell finite, namely the one in which the symmetry is enhanced to the full group of diffeomorphisms, i.e. Einstein's gravity, and another one denoted by WTDiff which enjoys local Weyl invariance. Both of them are free from the additional scalar. 
\end{abstract}

\end{titlepage}

\hrule
\tableofcontents
\vspace*{2em}
\hrule
\vspace*{2em}

\section{Introduction}
There are interesting theories of gravitation, such as Einstein's 1919
traceless one, which have the peculiarity that in order to obtain the field equations
from an action principle, the allowed variations $\d^t g_{\m\n}$ must obey $g^{\m\n}\d^t g_{\m\n}=0$. This particular model has attracted some attention since it provides a different point of view on the cosmological constant problem, in the sense that from Einstein's traceless equations one can obtain the usual equations of General Relativity with the addition of an arbitrary integration constant that plays the role of the cosmological constant (in \cite{Alvarez}
a general reference can be found).   
\par
A related slightly more drastic possibility is to restrict the symmetry of nature to the subgroup including those diffeomorphisms that enjoy unit Jacobian determinant, which we shall dub {\em transverse}\footnote{We are of course aware
of the fact that not every diffeomorphism close to the identity lies on a oner-parameter subgroup \cite{DeWitt}.} or TDiff. This subgroup is interesting because if the spacetime symmetry is TDiff, then nature makes no distinctions between tensor densities: they all transform as true tensors. In particular, the determinant of the metric is a scalar under TDiff, and we have freedom to include operators in our actions exactly as one does with usual matter scalars. In a related vein, we have recently proposed toy models in which not only the vacuum energy, but every form of potential energy does not gravitate \cite{Alvarezfaedo}, although there may be some subtleties in the coupling between matter and gravity \cite{Comment}. Transverse higher spin theories have been analyzed  in \cite{Skvortsov}.
\par
The aim of this work is to study transverse gravity, that is, gravity models enjoying this restricted symmetry principle, to one-loop order. For one reason: it has been shown that in many cases transverse theories are equivalent, at the classical level, to ordinary gravity with an extra scalar particle in the field content, akin to the usual dilaton, and with an arbitrary cosmological constant. This has been studied in some detail in the second reference in \cite{Alvarez}. As far as we know, quantum effects may spoil the equivalence. Furthermore, there are grounds to suspect that transverse theories ought to enjoy better ultraviolet behavior than Einstein's gravity, because there is no divergence associated to the conformal mode. On the other hand, there is no reason why this should be seen in perturbation theory. We shall begin by discussing the peculiarities of this kind of models with respect to General Relativity (GR) and the corresponding problem of performing a calculation where one cannot define a simple covariant gauge-fixing, and then we will study an equivalent scalar-tensor theory which is the one we eventually compute to one loop order.

\section{The one-loop transverse effective action}

The principal hypothesis we will adopt in this paper is that the spacetime symmetry of nature is not the full set of arbitrary coordinate changes, in the sense that it is assumed in Einstein's General Relativity, but only the subgroup of diffeomorphisms such that the determinant of the corresponding Jacobian equals unity. Once we assume this symmetry principle, a couple of important differences with respect to GR arise. 
\par
The first one of course is that now one is not able to distinguish between tensor densities of different weight. A tensor density is an object that under an active  change of coordinates (Diff)
\be
x^\m\,\to\, y^\m(x)
\ee
transforms as
\be
{T'}_{\m_1\dots\m_n}^{\n_1\dots\n_l}(y)=[D(y,x)]^\omega\,\,\frac{\pd x^{\r_1}}{\pd y^{\m_1}}\dots\frac{\pd x^{\r_n}}{\pd y^{\m_n}}\,\,\frac{\pd y^{\n_1}}{\pd x^{\s_1}}\dots\frac{\pd y^{\n_l}}{\pd x^{\s_l}}\,\,T_{\r_1\dots\r_n}^{\s_1\dots\s_l}(x)
\ee
where $\omega$ is the weight of the density and we have denoted
\be\label{jacob}
D(y,x)\equiv\det\left(\frac{\pd y^\m}{\pd x^\n}\right)
\ee
i.e.,  the determinant of the Jacobian. It is plain that were we to restrict our transformations to those that obey
\be\label{gauge}
D(y,x)=1,
\ee
 then every density behaves as a tensor. The most important consequence of this asumption is that two crucial scalar densities of GR, the determinant of the metric that represents the dynamics of gravity, as well as the integration element $d^nx$, are now a true scalar and  dual to a true scalar respectively. Therefore, we are free to use the determinant of the metric in the same way as any other scalar in the theory, writing down operators that were forbidden by the symmetry before.
 \par
  It has been shown in the second reference of \cite{Alvarez} that at the linear level models invariant under (the linealization of) TDiff propagate an additional degree of freedom included in the metric besides the usual spin two graviton. Eventually this mode will be responsible for an important piece of the divergences.
\par
Secondly, in GR arbitrary changes of coordinates are considered as a gauge symmetry that, as usual, one must fix. On a manifold of dimension $n$, there are $n$ gauge conditions one should give to gauge fix the local symmetry. Then, there is  enough room to do the fixing in a covariant way\footnote{Here and in what follows we are refering to covariance with respect to the background symmetry maintained in the Background Field method, and under which (\ref{harmgauge}) is a vector. As it is well known, the gauge fixing term must break the quantum symmetry (\ref{quansym}).}, which is very useful to simplify computations. An example of gauge commonly used is the harmonic (or minimal or DeWitt) gauge
\be\label{harmgauge}
\chi_\n\equiv\bn^\m h_{\m\n}-\frac{1}{2}\bn_\n h=0
\ee
where $h_{\m\n}$ is the graviton fluctuation, $h$ its trace with respect to the background metric and the covariant derivatives are constructed with the same background metric. Now, a slightly smaller symmetry means also less gauge conditions to fix. In particular, (\ref{gauge}) forces one of the $n$ original gauge parameters to be determined in terms of the others in such a way that there are only $n-1$ gauge conditions to specify, resulting in the impossibility to reach a convenient covariant gauge like the previous one. Of course, it is always possible to find, instead of a vector that vanishes and gives us the desired $n$ conditions, $n-1$ scalars constructed out of the graviton fluctuation, its trace and derivatives like for example
\be
\chi_1\equiv\bn^\m\bn^\n h_{\m\n}\hspace*{3em},\hspace*{3em}\chi_2\equiv\bn^2h\hspace*{3em},\hspace*{3em}\dots
\ee
The vanishing of these $\chi_1,\dots,\chi_{n-1}$ scalars constitutes an acceptable collection of gauge conditions to fix the TDiff symmetry of the system. Another possibility mentioned in \cite{Alvarez} is to project the harmonic gauge into the transverse direction
\be
\chi^t_\n\equiv\bn_\n\bn^\r\bn^\s h_{\r\s}-\bn^2\bn^\m h_{\m\n}
\ee
giving automatically $n-1$ independent conditions. Both gauge fixing choices, even if perfectly valid from a gauge theory point of view, are not suitable to undertake a calculation, the reason being that in general the operator obtained for the graviton fluctuations (and incidentally for the ghosts) does not take a minimal form, in a very precise sense. In particular, it cannot be put in the form of a Laplacian (see (\ref{oper})). Everyone that has worked out  a one-loop computation using Background Field methods and Heat Kernel techniques may appreciate the difficulties in dealing with non minimal operators, though there are known tractable examples \cite{Barvinskyvilkovisky}. 
\par
To avoid this unnecessary complication we will introduce a compensator field (some sort of Stueckelberg field) that renders the theory Diff invariant and so that we recover the original model in the, so to say, ``unitary gauge" (in analogy with the breaking of Electroweak symmetry) in which the compensator disappears from the spectrum. Imposing this partial gauge is one of the $n$ conditions of the Diff invariance and we are left with the $n-1$ conditions of TDiff, as it should be. The trick lies in maintaining the full invariance during the calculation in order to obtain a minimal operator fixing the gauge as in standard GR, but the price to pay is that the compensator will not vanish, since we have no symmetry left to reach the unitary gauge, and will be present in the final result. 
\par
Let us start with a particular example of a TDiff action, which is not the most general one can give. Consider the action
\be\label{action}
S_g=-\frac{1}{2\kappa^2}\,\,\int\,d^nx\,\,\sqrt{g^*}\,\,\left[f(g^*)\,R^*+2f_\l(g^*)\,\Lambda\right]
\ee
where $\Lambda$ plays the role of a cosmological constant, $f$ and $f_\l$ are arbitrary functions of the determinant of the metric $g^*\equiv\det{g^*_{\m\n}}$, and the action is in general not Diff invariant, except in the trivial case in which $f$ and $f_\l$ are constants. Moreover, under a Diff the action transforms to
\be
S_g=-\frac{1}{2\kappa^2}\,\,\int\,d^nx\,\,\sqrt{g^*}\,\,\left[f(g^*C^2)\,R^*+2f_\l(g^*C^2)\,\Lambda\right]
\ee
where $C(x)$ is a compensator field, defined so that $\varphi^*\equiv g^*C^2$ transforms as a true scalar (see (\ref{jacob})). The aforementioned unitary gauge would correspond to the choice $C=1$, recovering the original action. We can write in terms of the scalar field a perfectly Diff invariant action
\be
S_g=-\frac{1}{2\kappa^2}\,\,\int\,d^nx\,\,\sqrt{g^*}\,\,\left[f(\varphi^*)\,R^*+2f_\l(\varphi^*)\,\Lambda\right]
\ee
To perform the computation is convenient to go to the Einstein frame, so we make a conformal transformation
\bea
g_{\m\n}&=&\Omega^2 g_{\m\n}^*\nonumber\\
g&=&\Omega^{2n} g^*\nonumber\\
\varphi&=&gC^2=\Omega^{2n} g^*C^2=\Omega^{2n}\varphi^*
\eea
If we choose the conformal factor (supposing $n\neq 2$) as
\be\label{conf}
\Omega^{n-2}=f(\varphi^*)=f(\Omega^{-2n}\varphi)
\ee
then in terms of the new metric the action takes the form
\be
S_g=-\frac{1}{2\kappa^2}\,\,\int\,d^nx\,\,\sqrt{g}\,\,\left[R+2F_\l(\Omega)\,\Lambda\right]+\frac{(n-1)(n-2)}{2\kappa^2}\,\,\int\,d^nx\,\,\sqrt{g}\,\,\frac{1}{\Omega^2}\,\,g^{\m\n}\pd_\m\Omega \pd_\n\Omega
\ee
where we have made use of (\ref{conf}) in order to express $f_\l$ in terms of $\Omega$
\be
\Omega^{-n}f_\l(\Omega^{-2n}\varphi(\Omega))\equiv F_\l(\Omega)
\ee
Notice however that this reasoning cannot be applied when $f(g^*)=g^{*\frac{2-n}{2n}}$ since in that case
\be
\int\,d^nx\,\,\sqrt{g^*}\,f(\varphi^*)\,R^*=\int\,d^nx\,\,\sqrt{g}\,f(\varphi)\,R
\ee
	and one cannot get to the Einstein frame. A similar problem arises if $f(g^*)=\mathrm{constant}$ since then (\ref{conf}) is not invertible to give $\varphi^*=f^{-1}(\Omega^{n-2}$) or, in other words, we are already starting in the Einstein frame and the conformal transformation is not defined. Appart from these subtleties, a final redefinition of the scalar 
\be
\phi\equiv\sqrt{2(n-1)(n-2)}\,\,\ln{\Omega}
\ee
gives us the desired action
\be\label{suitact}
S_g=-\frac{1}{2\kappa^2}\,\,\int\,d^nx\,\,\sqrt{g}\,\,\left[R+2F_\l(\phi)\,\Lambda\right]+\frac{1}{2\kappa^2}\,\,\int\,d^nx\,\,\sqrt{g}\,\,\frac{1}{2}\,\,g^{\m\n}\pd_\m\phi \pd_\n\phi.
\ee
We have maintained the notation $F_\l(\phi)$ for $F_\l\left(\Omega\left(\phi\right)\right)$ in the hope it will cause
no confusion.
\par
In this form of the action the additional scalar degree of freedom is manifest, and it is suitable for performing the calculation using well known standard Background Field methods that, though straightforward, are quite tedious. The heavy details of the computation have been relegated to an appendix. The final result for the one-loop counterterm of the theory (\ref{action}), in terms of the original variables, reads
\bea
\Delta S&=&\frac{1}{\e}\,\frac{1}{ (4\pi)^2 }~\int\,d^4 x ~\sqrt{g^*}~\left\{\frac{1827}{160}\,f^{-4}\,f'^4\,\left(g_*^{\m\n}\pd_\m\varphi^*\pd_\n\varphi^*\right)^2 \right. \nonumber\\
&&\left.+\frac{171}{20}\,\Lambda\,f^{-3}\,f'^2\,f_\l\,g_*^{\m\n}\pd_\m\varphi^*\pd_\n\varphi^*-\frac{57}{5}\,\Lambda^2\, f^{-2}\,f_\l^2+\frac{1}{9}\,\Lambda^2\,\left[f'^{-1}\,f'_\l-2f^{-1}\,f_\l\right]^2\right.\nonumber\\
&&\left.+\frac{2}{9}\,\Lambda^2\,f^2\left[4\,f^{-2}\,f_\l-3f^{-1}\,f'^{-1}\,f'_\l+f'^{-2}\,f''_\l-f'^{-3}\,f''\,f'_\l\right]\right.\nonumber\\
&&\left.\times\left[2\,f^{-2}\,f_\l-3f^{-1}\,f'^{-1}\,f'_\l+f'^{-2}\,f''_\l-f'^{-3}\,f''\,f'_\l\right]\right.\bigg\}
\eea
Prime denotes derivative with respect to $\varphi^*$. It is clear that our cherished hope that, in the absence of the conformal mode, the ultraviolet behavior of transverse models could be better than the correspondent in GR is not fulfilled. Even if the cosmological constant vanishes, the first term in the counterterm remains, except in case $f$ is constant, but that corresponds exactly to the Einstein--Hilbert action, which is known to be one-loop finite \cite{Hooft}. In fact, the form of the counterterm reminds the one obtained when a scalar field, possibly with a potential term, is coupled to gravity. That is because the mode responsible for the divergences is the additional mode in the metric which cannot be killed in the lack of full Diff invariance, as will become more transparent in what follows.  
\par
We should  mention here that the GR limit $f'\to 0$ is not regular if the cosmological constant does not vanish. But remember that this limit is one of the problematic cases regarding the conformal transformation (see the coments following (\ref{conftrans})). Also, quantum corrections tend to generate a kinetic energy term for the determinant of the metric, so it is convenient to include it in the bare action from the beginning, obtaining a more complete model.

\section{A  more general transverse action}

Taking into account the last considerations, in this section we will extend the model by introducing a kinetic energy term for the determinant of the metric. The resulting action will be the most general gravitatory TDiff action with the usual properties one imposes to a suitable action (to be a scalar of the symmetry, second order in derivatives etc.)\footnote{One could have included a potential term
\be
S_V=-\frac{1}{2\kappa^2}\,\int d^nx\,\,\sqrt{g^*}\,M^2\,V(g^*)
\ee
but it can be absorbed in the definition of $F_\l(\Omega)$, i.e.,
\be
2\Lambda F_\l(\Omega)\equiv\Omega^{-n}\left(2\Lambda f_\l(f^{-1}(\Omega^{n-2})+M^2V(f^{-1}(\Omega^{n-2})\right)
\ee
so it does not include any interesting new issue and we won't consider it.}
\be\label{kiner}
S=-\frac{1}{2\kappa^2}\,\int d^nx\,\,\sqrt{g^*}\,\,\left[ f(g^*)R^*+2f_\l(g^*)\Lambda+\frac{1}{2}f_\phi(g^*)g_*^{\m\n}\pd_\m g^*\pd_\n g^*\right]
\ee
so that, as before, after an arbitrary change of coordinates
\be
S=-\frac{1}{2\kappa^2}\,\int d^nx\,\,\sqrt{g^*}\,\,\left[ f(\varphi^*)R^*+2f_\l(\varphi^*)\Lambda+\frac{1}{2}f_\phi(\varphi^*)g_*^{\m\n}\pd_\m \varphi^*\pd_\n \varphi^*\right]
\ee
where the scalar field is $\varphi^*\equiv g^*C^2$. We should now go to the Einstein frame through a conformal transformation
\be
g_{\m\n}=\Omega^2g^*_{\m\n}
\ee
Choosing the conformal factor as
\be\label{conftrans}
\Omega^{n-2}=f(\varphi^*)
\ee
the action in the new frame takes the form
\bea
S&=&-\frac{1}{2\kappa^2}\,\int d^nx\,\,\sqrt{g}\,\,\left[R+2F_\l(\Omega)\Lambda\right]+\frac{1}{2\kappa^2}\,\int d^nx\,\,\sqrt{g}\,\,\left[\frac{2(n-1)(n-2)}{\Omega^2}\right.\nonumber\\
&&\left.-\Omega^{2-n}\,f_\phi\left(f^{-1}(\Omega^{n-2})\right)\left(\frac{\pd f^{-1}(\Omega^{n-2})}{\pd \Omega}\right)^2\right]\,\frac{1}{2}\,g^{\m\n}\pd_\m\Omega\pd_\n\Omega
\eea
where we have defined
\be\label{potential}
F_\l(\Omega)\equiv\Omega^{-n}\,f_\l\left(f^{-1}(\Omega^{n-2})\right)
\ee
A final redefinition of the scalar gives the desired action studied earlier
\be\label{defsca}
\left[\frac{2(n-1)(n-2)}{\Omega^2}-\Omega^{2-n}\,f_\phi\left(f^{-1}(\Omega^{n-2})\right)\left(\frac{\pd f^{-1}(\Omega^{n-2})}{\pd \Omega}\right)^2\right]\,g^{\m\n}\pd_\m\Omega\pd_\n\Omega=g^{\m\n}\pd_\m\phi\pd_\n\phi
\ee
and consequently we can use the counterterm quoted in the appendix. We have fixed the sign of the kinetic term of the new field $\phi$ so that it is not a ghost, and then we are forced to require the function of the left hand side to be positive definite. In terms of the original functions it means
\be\label{bound}
2(n-1)\,f^{\frac{2}{2-n}}-(n-2)\,f_\phi\,f^{\frac{n-4}{n-2}}\,f'^{-2}\geq0
\ee
Finally, we are able to write the one-loop counterterm of the theory (\ref{kiner})
\bea
\Delta S&=&\frac{1}{\e}\frac{1}{ (4\pi)^2 }~\int\,d^4 x ~\sqrt{g^*} ~ \left\{\frac{203}{160}\left[3f^{-2}\,f'^2-f^{-1}f_\phi\right]^2\left(g_*^{\m\n}\pd_\m\varphi^*\pd_\n\varphi^*\right)^2\right.\nonumber\\
&&\left.+\frac{57}{20}\,\Lambda\left[3f^{-3}\,f'^2\,f_\l-f^{-2}\,f_\l\,f_\phi\right]g_*^{\m\n}\pd_\m\varphi^*\pd_\n\varphi^*-\frac{57}{5}\,\Lambda^2\,f^{-2}\,f_\l^2\right.\nonumber\\
&&\left.+\frac{1}{3}\,\Lambda^2\left[f'^{-1}\,f'_\l-2f^{-1}\,f_\l\right]^2\left[3-f\,f'^{-2}\,f_\phi\right]^{-1}+\frac{1}{2}\,\Lambda^2\left[3f^{-1}-f'^{-2}\,f_\phi\right]^{-4}\right.\nonumber\\
&&\left.\times\left[24f^{-3}\,f_\l-18f^{-2}\,f'^{-1}\,f'_\l-6f^{-1}\,f'^{-3}\,f''\,f'_\l+6f^{-1}\,f'^{-2}\,f''_\l-10f^{-2}\,f'^{-2}\,f_\l\,f_\phi\right.\right.\nonumber\\
&&\left.\left.+7f^{-1}\,f'^{-3}\,f'_\l\,f_\phi-2f^{-1}\,f'^{-3}\,f_\l\,f'_\phi+4f^{-1}\,f'^{-4}\,f''\,f_\l\,f_\phi-2f'^{-4}\,f''_\l\,f_\phi+f'^{-4}\,f'_\l\,f'_\phi\right]\right.\nonumber\\
&&\left.\times\left[12f^{-3}\,f_\l-18f^{-2}\,f'^{-1}\,f'_\l-6f^{-1}\,f'^{-3}\,f''\,f'_\l+6f^{-1}\,f'^{-2}\,f''_\l-2f^{-2}\,f'^{-2}\,f_\l\,f_\phi\right.\right.\nonumber\\
&&\left.\left.+7f^{-1}\,f'^{-3}\,f'_\l\,f_\phi-2f^{-1}\,f'^{-3}\,f_\l\,f'_\phi+4f^{-1}\,f'^{-4}\,f''\,f_\l\,f_\phi-2f'^{-4}\,f''_\l\,f_\phi+f'^{-4}\,f'_\l\,f'_\phi\right.\right.\nonumber\\
&&\left.\left.-\frac{4}{3}f^{-1}\,f'^{-4}\,f_\l\,f_\phi^2\right]
\right\}
\eea
Let us remark that when  $\Lambda=0$ and the functions in front of the kinetic term and the Einstein--Hilbert term are $f=f_\phi=1$ we recover the result of 't Hooft and Veltman for gravity coupled to a scalar without potential
\bea
\Delta S&=&\frac{1}{\e}\frac{1}{ (4\pi)^2 }~\int\,d^4 x ~\sqrt{g^*} ~ \frac{203}{160}\left(g_*^{\m\n}\pd_\m\varphi^*\pd_\n\varphi^*\right)^2\nonumber\\
&=&\frac{1}{\e}\frac{1}{ (4\pi)^2 }~\int\,d^4 x ~\sqrt{g^*} ~ \frac{203}{40}\,R^{*2}
\eea
\par
Notice in passing that now the limit $f'\to 0$ is not singular, and this is due to the presence of a kinetic energy term for the scalar even if the conformal transformation is not defined. Moreover, the following diferential equation relating both functions
\be\label{difeq}
2(n-1)f^{-1}\,f'^2-(n-2)f_\phi=0
\ee
saturates the bound (\ref{bound}) and has a real solution if the product $f\,f_\phi$, as a function of the determinant of the metric, is positive definite, and therefore there is another family of one-loop finite theories in case $\Lambda=0$. Nevertheless, after an easy computation one may prove that given the action (\ref{kiner}), and under the hypothesis that the arbitrary functions verify (\ref{difeq}), it (almost) always exists a conformal transformation, which is preciselly (\ref{conftrans}), that leads to the Einstein--Hilbert action. As a result, the family of theories (\ref{difeq}) are nothing but GR written in another frame, with full Diff invariance. Under this point of view the one-loop finiteness is not surprising. 
\par
The only theory that cannot be put in Einstein--Hilbert form is precisely when 
\be
f(g^*)=g^{*\frac{2-n}{2n}}
\ee
and $f_\phi$ is given by (\ref{difeq}). It can be seen that this theory has an additional local Weyl symmetry
\be
g_{\m\n}\to\Omega^2(x)g_{\m\n}
\ee
that prevents us from going to the Einstein frame. In fact, this theory is exactly the WTDiff model of the second reference in \cite{Alvarez}, which is a  ``unimodular" model (that is, a theory that can be written in terms of a metric with unit determinant and nothing else) but written in terms of a metric not restricted to have unit determinant. 
\par
It is important to mention that both finite transverse theories have an enhanced symmetry that allows us to remove from the spectrum the additional degree of freedom contained in the metric. Then, as we have repeatedly advertised, the observed worse behavior in the ultraviolet is due to this mode.

\section{Conclusions}
The main conclusion of our investigation is that there are only two transverse theories of gravity that are finite
on shell.  The first one appears when TDiff is enhanced to Diff (and besides, the cosmological constant is fine tuned to zero); that is Einstein's gravity, whose on shell one-loop finiteness was proven in a classic work by 't Hooft and Veltman \cite{Hooft}.
\par  
The other theory enjoys also a greater symmetry, a local Weyl invariance denoted WTDiff, that allows to remove the additional degree of freedom present in generic transverse models. One could then be sure that the divergence found is precisely due to this mode. WTDiff theories include the so-called {\em unimodular} ones, which  can be written using only the metric $\hat{g}_{\m\n}$ such that $\det\,\hat{g}_{\m\n}=1$, but the class of WTDiff could perhaps be larger that the unimodular one. Here we should mention that the computation was done in the Einstein frame, in which there is no function (other than the square root of the determinant) in front of the curvature scalar. However, as we have said, technically is not possible to reach this frame in a theory with Weyl invariance like WTDiff, for obvious reasons. Therefore, though the model with WTDiff verifies (\ref{difeq}), it falls into the cases we cannot treat with our formalism. Strictly speaking one should then repeat the calculation in an arbitrary frame to be sure of the conclusions, but at the end the result will certanly be the same since its physical origin seems to be clear: the absence of the scalar mode. 
\par
It should be remarked also that we actually have calculated in a Diff invariant theory which coincides with the transverse theory of our interest in the unitary gauge $C=1$. Our computation was done in the equivalent of the renormalizable gauge for Yang--Mills theories, and it does ultimately rely on gauge invariance of the extended theory. In this sense it would be interesting to extend the analysis of the existence of a nilpotent BRS symmetry perhaps along the lines of what was done for transverse theories in \cite{Dragon}.

\section*{Acknowledgments}
We are grateful for useful discussions and/or correspondence with Drs D.~Blas and A.~Y.~Kamenshchik. This work has been partially supported by the European Commission (HPRN-CT-200-00148) and by FPA2006-05423 (DGI del MCyT, Spain), Proyecto HEPHACOS ; P-ESP-00346 (CAM) and Consolider PAU, CSD-2007-00060. A.F.F. has been supported by a MEC grant, AP-2004-0921. J.J.L-V. wants to thank J.~M.~Mart\'in-Garc\'ia
for his help regarding the use of the package xTensor \cite{Martin}, which was used to check the algebraic computations.    

\appendix
\section{Some details of the computations}
To begin, let us be quite explicit on our notation and conventions. 
\begin{itemize}
\item
The flat tangent metric is mostly negative
\be
\eta_{ab}\equiv diag\,\left(1,-1,-1,-1\right).
\ee
The Riemann tensor is
\be
R^\m\,_{\n\a\b}\equiv \pd_\a \Gamma^\m_{\n\b}-\pd_\b \Gamma^\m_{\n\a}+
\Gamma^\m_{\sigma\a}\Gamma^{\sigma}_{\n\b}-\Gamma^\m_{\sigma\b}\Gamma^\sigma_{\n\a}
\ee
and we define the Ricci tensor as
\be
R_{\m\n}\equiv R^\l\,_{\m\l\n}.
\ee
Our conventions for the cosmological constant are such that for a constant curvature space
\be
R_{\m\n}=-\frac{2}{n-2}\l g_{\m\n}
\ee
then the ordinary de Sitter space  has negative constant curvature, but enjoys {\em positive cosmological constant}. The Einstein--Hilbert action is consequently defined as
\be
S=-\frac{c^3}{2\kappa^2}\int d^nx\sqrt{|g|}\left(R+2\lambda\right)+S_{matter}
\ee
with $\kappa^2\equiv 8\pi G$.
\item
Background covariant derivatives can be integrated by parts:
\be
\int d^n x \sqrt{|\bar{g}|}\,\bar{\nabla}_\m L^\m=\int d^n x \sqrt{|\bar{g}|}
\,\frac{1}{\sqrt{|\bar{g}|}}\,\pd_\m \left(\sqrt{|\bar{g}|}L^\m\right)=\int d^n x \,\pd_\m \left(\sqrt{|\bar{g}|}L^\m\right)
\ee
and some useful commutators with our conventios are:
\bea
\left[\bn_\b,\bn_\gamma\right]\omega_\rho&=&\omega_\m \bar{R}^\m\,_{\rho\g\b}\nonumber\\
\left[\bn_\b,\bn_\gamma\right]V^\rho&=&-V^\m \bar{R}^\rho\,_{\m\g\b}\nonumber\\
\left[\bn_\b,\bn_\gamma\right]h^{\a\b}&=&-h^{\l\b} \bar{R}^\a\,_{\l\gamma\b}+h^{\a\l} 
\bar{R}_{\l\gamma}
\eea

\item Let us now begin with the analysis proper, pointing out only the different steps of the calculation. Both the metric and the scalar field in the action (\ref{suitact}) are expanded in a background field and a perturbation
\bea
g_{\m\n}&=&\bg_{\m\n}+\kappa\, h_{\m\n}\nonumber\\
g^{\m\n}&=&\bg^{\m\n}-\kappa\,h^{\m\n}+\kappa^2\,h^\m{}_\a h^{\a\n}+O(\kappa^3)\nonumber\\
\phi&=&\bp+\kappa\,\phi.
\eea
Where indices are raised with the background metric and geometric quantities (curvature tensors, covariant derivatives...) calculated with respect to this metric wear a bar. To take into account one-loop effects it is enough to expand the action up to quadratic order in the perturbations. After expanding, the term linear in the coupling cancels due to the background equations of motion, namely
\bea\label{beom}
\bn^2\bp+2\Lambda F'_\l(\bp)&=&0\nonumber\\
\br_{\m\n}-\frac{1}{2}\br\bg_{\m\n}-\Lambda F_\l(\bp)\bg_{\m\n}-\frac{1}{2}\bn_\m\bp\bn_\n\bp+\frac{1}{4}\bg_{\m\n}\bg^{\a\b}\bn_\a\bp\bn_\b\bp&=&0
\eea
and prime denotes derivative with respect to $\phi$. Using the known expansion for the scalar curvature the quadratic order operator is
\bea
S_g&=&\frac{1}{2}\,\,\int\,d^nx\,\,\sqrt{\bg}\,\,\left[h^{\a\b}\left(\frac{1}{4}\bg_{\a\b}\bg_{\m\n}\bn^2-\frac{1}{4}\bg_{\a\m}\bg_{\b\n}\bn^2+\frac{1}{2}\bg_{\a\m}\bn_\b\bn_\n-\frac{1}{2}\bg_{\m\n}\bn_\a\bn_\b\right.\right.\nonumber\\
  &&\left.\left.+\frac{1}{2}\bg_{\a\b}\br_{\m\n}-\frac{1}{2}\bg_{\a\m}\br_{\b\n}-\frac{1}{2}\br_{\a\m\b\n}+\frac{1}{2}\bg_{\a\m}\pd_\b\bp\pd_\n\bp-\frac{1}{4}\bg_{\a\b}\pd_\m\bp\pd_\n\bp\right.\right.\nonumber\\
&&\left.\left.-\frac{\left(\br+2\Lambda F_\l(\bp)-\frac{1}{2}\bg^{\r\s}\pd_\r\bp\pd_\s\bp\right)}{8}\left(\bg_{\a\b}\bg_{\m\n}-2\bg_{\a\m}\bg_{\b\n}\right)\right)h^{\m\n}\right.\nonumber\\
&&\left.+h^{\a\b}\left(\frac{1}{2}\bg_{\a\b}\bg^{\r\s}\pd_\r\bp\pd_\s-\pd_\a\bp\pd_\b-\Lambda\bg_{\a\b}F'_\l(\bp)\right)\phi+\phi\left(-\frac{1}{2}\bn^2-\Lambda F^{''}_\l(\bp)\right)\phi\right]\nonumber\\
\eea
At this stage the operator is very cumbersome, but we still have the freedom to fix the gauge in a way that simplifies the computation, since we have been careful enough to include the compensator to increase the symmetry to full Diff. Taking the expresion
\be
\chi_\n=\bn^\m h_{\m\n}-\frac{1}{2}\bn_\n h-\phi\pd_\n\bp
\ee
we choose as gauge fixing term
\be
S_{gf}=\frac{1}{2}\,\,\int\,d^nx\,\,\sqrt{\bg}\,\,\frac{1}{2\xi}\,\,\bg^{\m\n}\chi_\m\chi_\n
\ee
which after expanding can be expressed in the form
\bea
S_{gf}&=&\frac{1}{2}\,\,\int\,d^nx\,\,\sqrt{\bg}\,\,\frac{1}{2\xi}\,\,\left[h^{\a\b}\left(\bg_{\m\n}\bn_\a\bn_\b-\bg_{\a\m}\bn_\b\bn_\n-\frac{1}{4}\bg_{\a\b}\bg_{\m\n}\bn^2\right)h^{\m\n}\right.\nonumber\\
&&\left.+2h^{\a\b}\left(\pd_\a\bp\pd_\b+\bn_\a\bn_\b\bp-\frac{1}{2}\bg_{\a\b}\bg^{\r\s}\pd_\r\bp\pd_\s-\frac{1}{2}\bg_{\a\b}\bg^{\r\s}\bn_\r\bn_\s\bp\right)\phi\right.\nonumber\\
&&\left. +\phi\left(\bg^{\a\b}\pd_\a\bp\pd_\b\bp\right)\phi\right]
\eea
Let us define the following tensor with the desired symmetry properties, i.e., symmetric in $(\m\n)$, $(\a\b)$ and under the interchange $(\m\n)\leftrightarrow (\a\b)$
\bea
C_{\a\b\m\n}&=&\frac{1}{4}\left(\bg_{\a\m}\bg_{\b\n}+\bg_{\a\n}\bg_{\b\m}-\bg_{\a\b}\bg_{\m\n}\right)\nonumber\\
C^{\a\b\m\n}&=&\bg^{\a\m}\bg^{\b\n}+\bg^{\a\n}\bg^{\b\m}-\frac{2}{n-2}\bg^{\a\b}\bg^{\m\n}\nonumber\\
\d^{\a\b}_{\m\n}&=&\d^{(\a}_\m\d^{\b)}_\n
\eea
the full action can be written as
\be
S_g+S_{gf}=\frac{1}{2}\,\,\int\,d^nx\,\,\sqrt{\bg}\,\,\frac{1}{2}\,\,\left[h^{\a\b}M_{\a\b\m\n}h^{\m\n}+h^{\a\b}D_{\a\b}\phi+\phi E_{\m\n}h^{\m\n}+\phi F\phi\right]
\ee
where the operators are 
\bea
M_{\a\b\m\n}&=&C_{\a\b\r\s}\left(-\d^{\r\s}_{\m\n}\bn^2+\frac{1-\xi}{\xi}\bg_{\m\n}\bn^{(\r}\bn^{\s)}+\frac{2(\xi-1)}{\xi}\d^{(\r}_{(\m}\bn^{\s)}\bn_{\n)}+P_{\m\n}^{\r\s}\right)\nonumber\\
P_{\m\n}^{\r\s}&=&-2\br^{(\r}{}_\m{}^{\s)}{}_\n-2\d^{(\r}_{(\m}\bar{R}^{\s)}_{\n)}+\left(\bar{R}+2\Lambda F_\l(\bp)-\frac{1}{2}\bg^{\a\b}\pd_\a\bp\pd_\b\bp\right)\d^{\r\s}_{\m\n}+\bg^{\r\s}\br_{\m\n}\nonumber\\
&&+\frac{2}{(n-2)}\bg_{\m\n}\br^{\r\s}-\frac{1}{(n-2)}\bg_{\m\n}\bg^{\r\s}\br+2\d^{(\r}_{(\m}\pd_{\n)}\bp\pd^{\s)}\bp-\frac{1}{2}\bg_{\m\n}\pd^\r\bp\pd^\s\bp\nonumber\\
&&-\frac{1}{(n-2)}\bg^{\r\s}\pd_\m\bp\pd_\n\bp+\frac{1}{2(n-2)}\bg_{\m\n}\bg^{\r\s}\pd_\l\bp\pd^\l\bp\nonumber\\
D_{\a\b}&=&\frac{2(1-\xi)}{\xi}\,C_{\a\b\r\s}\,\bn^\r\bp\bn^\s+\frac{\xi+1}{\xi}\,C_{\a\b\r\s}\bn^\r\bn^\s\bp-\Lambda F'_\l(\bp)\bg_{\a\b}\nonumber\\
E_{\m\n}&=&\frac{2(\xi-1)}{\xi}\,C_{\m\n\r\s}\,\bn^\r\bp\bn^\s+\frac{\xi+1}{\xi}\,C_{\m\n\r\s}\bn^\r\bn^\s\bp-\Lambda F'_\l(\bp)\bg_{\m\n}\nonumber\\
F&=&-\bn^2-2\Lambda F^{''}_\l(\bp)+\frac{1}{\xi}\bg^{\r\s}\pd_\r\bp\pd_\s\bp
\eea
in such a way that in terms of the combined field
\be
\psi^A\equiv\left(\begin{array}{c}
h^{\m\n}\\
\phi
\end{array}\right)
\ee
and in the minimal gauge, corresponding to $\xi=1$, the operator
\be
S=\frac{1}{2}\,\,\int\,d^nx\,\,\sqrt{\bg}\,\frac{1}{2}\,\psi^A\Delta_{AB}\psi^B
\ee
is minimal, in the sense that it takes a Laplacian form
\be\label{oper}
\Delta_{AB}=-g_{AB}\bn^2+Y_{AB}
\ee
with the metric
\be
g_{AB}=\left(\begin{array}{cc}
C_{\a\b\m\n} & 0 \\
0 & 1 \end{array} \right)
\ee
the inverse metric
\be
g^{AB}=\left(\begin{array}{cc}
C^{\a\b\m\n} & 0 \\
0 & 1 \end{array} \right)
\ee
and the term without derivatives
\be
Y_{AB}=\left(\begin{array}{cc}
C_{\a\b\r\s}P^{\r\s}_{\m\n} & 2C_{\a\b\r\s}\bn^\r\bn^\s\bp-\Lambda F'_\l(\bp)\bg_{\a\b} \\
2C_{\m\n\r\s}\bn^\r\bn^\s\bp-\Lambda F'_\l(\bp)\bg_{\m\n} & -2\Lambda F^{''}_\l(\bp)+\bg^{\r\s}\pd_\r\bp\pd_\s\bp
\end{array}\right)
\ee
\item On the other hand, once we have an operator in the Laplacian form (\ref{oper}), the one-loop counterterm (supposing that we work in $n=4$ dimensions) is given by the following coefficient in the heat kernel expansion \cite{Barvinskyvilkovisky}
\bea\label{coe}
a_4&=&\frac{1}{(4\pi)^{\frac{n}{2}}}\,\frac{1}{360}\,\int d^nx\,\sqrt{\bg}\,\,\tr\,\left(180Y^2-60\br Y+5\br^2-\right.\nonumber\\
&&\left.-2\br_{\m\n}\br^{\m\n}+2\br_{\m\n\r\s}\br^{\m\n\r\s}+30W_{\m\n}W^{\m\n}\right)
\eea
and the field strength is defined through
\be
[\bn_\m,\bn_\n]\psi^A=W_{B\m\n}^A\psi^B
\ee
Therefore, in order to find the counterterm we will need the following traces
\bea
\tr\, \mathbb{I}&=&\d^{\a\b}_{\a\b}+1=\frac{n(n+1)+2}{2}\nonumber\\
\tr\, Y&=&g^{AB}Y_{AB}=\d_{\a\b}^{\m\n}P_{\m\n}^{\a\b}-2\Lambda F^{''}_\l(\bp)+\bg^{\r\s}\pd_\r\bp\pd_\s\bp\nonumber\\
&=&\frac{n(n+1)}{2}\left(\bar{R}+2\Lambda F_\l(\bp)-\frac{1}{2}\bg^{\r\s}\pd_\r\bp\pd_\s\bp\right)-n\br+(n+2)\bg^{\r\s}\pd_\r\bp\pd_\s\bp-2\Lambda F^{''}_\l(\bp)\nonumber\\
\tr\, Y^2&=&Y_{AB}\,g^{BC}\,Y_{CD}\,g^{DA}=P^{\a\b}_{\m\n}P^{\m\n}_{\a\b}+2D_{\a\b}E_{\m\n}C^{\m\n\a\b}+\left(2\Lambda F^{''}_\l(\bp)-\bg^{\r\s}\pd_\r\bp\pd_\s\bp\right)^2\nonumber\\
&=&3\br_{\m\n\r\s}\br^{\m\n\r\s}+\frac{n^2-8n+4}{n-2}\br_{\m\n}\br^{\m\n}+\frac{n+2}{n-2}\br^2-2n\br\bigg(\bar{R}+2\Lambda F_\l(\bp)\nonumber\\
&&\left.-\frac{1}{2}\bg^{\r\s}\pd_\r\bp\pd_\s\bp\right)+\frac{n\left(n+1\right)}{2}\left(\bar{R}+2\Lambda F_\l(\bp)-\frac{1}{2}\bg^{\r\s}\pd_\r\bp\pd_\s\bp\right)^2+2\bn^2\bp\bn^2\bp\nonumber\\
&&-8\Lambda F'_\l\,\bn^2\bp-\frac{8n}{n-2}\Lambda^2(F'_\l(\bp))^2+\frac{n^2-5}{n-2}\left(\bg^{\r\s}\pd_\r\bp\pd_\s\bp\right)^2\nonumber\\
&&+\frac{n(4-n)(3n-8)-4(n-2)^2}{(n-2)^2}\,\br^{\m\n}\pd_\m\bp\pd_\n\bp-\frac{n^2+4n-16}{(n-2)^2}\,\br\,\bg^{\r\s}\pd_\r\bp\pd_\s\bp\nonumber\\
&&+2\left(n+1\right)\left(\br+2\Lambda F_\l(\bp)-\frac{1}{2}\bg^{\r\s}\pd_\r\bp\pd_\s\bp\right)\bg^{\g\l}\pd_\g\bp\pd_\l\bp+\left(2\Lambda F^{''}_\l(\bp)-\bg^{\r\s}\pd_\r\bp\pd_\s\bp\right)^2\nonumber\\
\tr\, W_{\m\n}W^{\m\n}&=&-(n+2)\br_{\m\n\r\s}\br^{\m\n\r\s}
\eea
\item Using the known expression (\ref{coe}) of the fourth heat kernel coefficient one gets
\bea
a_4 &=& \frac{1}{ (4\pi)^{\frac{n}{2}} } ~\frac{1}{360} ~\int\,d^n x ~\sqrt{\bg} ~\left\{ \left[ 542+n(n+1)-30(n+2) \right] \br_{\m\n\r\s}\br^{\m\n\r\s}\right.\nonumber\\
&&\left. +\left[ 180\frac{n^2-8n+4}{n-2}-n(n+1)-2 \right] \br_{\m\n}\br^{\m\n} +\left[ 180\frac{n+2}{n-2}+60n+\frac{5n(n+1)+10}{2} \right] \br^2\right.\nonumber\\
&&\left.-30n(n+13)\br \left(\bar{R}+2\Lambda F_\l(\bp)-\frac{1}{2}\bg^{\r\s}\pd_\r\bp\pd_\s\bp\right)+90n(n+1)\bigg(\bar{R}+2\Lambda F_\l(\bp)\right.\nonumber\\
&&\left.\left.-\frac{1}{2}\bg^{\r\s}\pd_\r\bp\pd_\s\bp\right)^2+180\frac{n(4-n)(3n-8)-4(n-2)^2}{(n-2)^2}\,\br^{\m\n}\pd_\m\bp\pd_\n\bp\right.\nonumber\\
&&\left.-60\left[3\frac{n^2+4n-16}{(n-2)^2}+(n+2)\right]\br\,\bg^{\r\s}\pd_\r\bp\pd_\s\bp+360\left(n+1\right)\bigg(\br+2\Lambda F_\l(\bp)\right.\nonumber\\
&&\left.\left.-\frac{1}{2}\bg^{\r\s}\pd_\r\bp\pd_\s\bp\right)\bg^{\g\l}\pd_\g\bp\pd_\l\bp+180\frac{n^2+n-7}{n-2}\left(\bg^{\r\s}\pd_\r\bp\pd_\s\bp\right)^2+360\bn^2\bp\bn^2\bp\right.\nonumber\\
&&\left.+120\Lambda\br F^{''}_\l(\bp)-1440\Lambda F'_\l(\bp)\bn^2\bp-720\Lambda F^{''}_\l(\bp)\bg^{\r\s}\pd_\r\bp\pd_\s\bp-\frac{1440n}{n-2}\Lambda^2(F'_\l(\bp))^2\right.\nonumber\\
&&\left.+720\Lambda^2(F^{''}_\l(\bp))^2\right\}
\eea
\item Remember that we need also the contribution coming from ghost loops. The gauge fixing term mantains background invariance, under which the background $\bar{g}_{\m\n}$ transforms as a metric and the fluctuation $h_{\m\n}$ as a tensor. On the other hand it has to break the quantum symmetry
\bea\label{quansym}
\d\bg_{\m\n}&=&0\nonumber\\
\d h_{\m\n}&=&\frac{2}{\kappa}\bn_{(\m}\xi_{\n)}+\mathcal{L}_\xi h_{\m\n}\nonumber\\
\d\bp&=&0\nonumber\\
\d \phi&=&\frac{1}{\kappa}\xi^\m\bn_\m\left(\bp+\kappa \phi\right)
\eea
The ghost Lagrangian is obtained performing a variation on the gauge fixing term 
\be
\d\chi_\n=\frac{1}{\kappa}\left(\bn^2\bg_{\m\n}+\br_{\m\n}-\bn_\m\bp\bn_\n\bp\right)\xi^\m
\ee
plus terms that give operators cubic in fluctuations and therefore are irrelevant at one loop (the ghosts are always quantum fields, they do not appear as external states). Then, as ghost Lagrangian we will take 
\be
S_{gh}=\frac{1}{2}\,\,\int\,d^nx\,\,\sqrt{\bg}\,\frac{1}{2}\,V^*_\m\left(-\bn^2\bg^{\m\n}-\br^{\m\n}+\bn^\m\bp\bn^\n\bp\right)V_\n
\ee
The relevant traces are
\bea
\tr\, \mathbb{I}&=& n\nonumber\\
\tr\,Y&=&-\br+\bg^{\r\s}\pd_\r\bp\pd_\s\bp\nonumber\\
\tr\,Y^2&=&\br_{\m\n}\br^{\m\n}-2\br^{\m\n}\pd_\m\bp\pd_\n\bp+\left(\bg^{\r\s}\pd_\r\bp\pd_\s\bp\right)^2\nonumber\\
\tr\, W_{\m\n}W^{\m\n}&=&-\br_{\m\n\r\s}\br^{\m\n\r\s}
\eea
and the coefficient
\bea
a^{gh}_4 &=& \frac{1}{ (4\pi)^{\frac{n}{2}} } ~\frac{1}{360} ~\int\,d^n x ~\sqrt{\bg} ~\left\{\left[2n-30\right]\br_{\m\n\r\s}\br^{\m\n\r\s}+\left[180-2n\right]\br_{\m\n}\br^{\m\n}\right.\nonumber\\
&&\left.+\left[60+5n\right]\br^2-360\br^{\m\n}\pd_\m\bp\pd_\n\bp-60\br\bg^{\r\s}\pd_\r\bp\pd_\s\bp+180\left(\bg^{\r\s}\pd_\r\bp\pd_\s\bp\right)^2\right\}\nonumber\\
\eea
\item Adding the two pieces together and particularizing to the physical dimension $n=4$ one gets the one-loop counterterm (notice the factor and the sign of the ghost contribution)
\bea\label{count}
\Delta S&=&\frac{1}{\e}\left(a_4-2a^{gh}_4\right)= \frac{1}{\e}\,\frac{1}{ (4\pi)^2 } ~\int\,d^4 x ~\sqrt{\bg} ~\left\{ \frac{71}{60} \br_{\m\n\r\s}\br^{\m\n\r\s}-\frac{241}{60}\br_{\m\n}\br^{\m\n} + \frac{15}{8} \br^2\right.\nonumber\\
&&\left.-\frac{17}{3}\br \left(\bar{R}+2\Lambda F_\l(\bp)-\frac{1}{2}\bg^{\r\s}\pd_\r\bp\pd_\s\bp\right)+5\left(\bar{R}+2\Lambda F_\l(\bp)-\frac{1}{2}\bg^{\r\s}\pd_\r\bp\pd_\s\bp\right)^2\right.\nonumber\\
&&\left.-\frac{8}{3}\br\,\bg^{\r\s}\pd_\r\bp\pd_\s\bp+5\left(\br+2\Lambda F_\l(\bp)-\frac{1}{2}\bg^{\r\s}\pd_\r\bp\pd_\s\bp\right)\bg^{\g\l}\pd_\g\bp\pd_\l\bp+\frac{9}{4}\left(\bg^{\r\s}\pd_\r\bp\pd_\s\bp\right)^2\right.\nonumber\\
&&\left.+\bn^2\bp\bn^2\bp+\frac{1}{3}\Lambda\br F^{''}_\l(\bp)-4\Lambda F'_\l(\bp)\bn^2\bp-2\Lambda F^{''}_\l(\bp)\bg^{\r\s}\pd_\r\bp\pd_\s\bp-8\Lambda^2(F'_\l(\bp))^2\right.\nonumber\\
&&\left.+2\Lambda^2(F^{''}_\l(\bp))^2
\right\}
\eea
After having obtained this result, we have found in the literature a completely equivalent computation \cite{Barvinsky}. The counterterm of Barvinsky {\em et al.} is expressed in a different gauge, but using the background equations of motion one can go from one to the other. This a consequence of the old theorem asserting that the pieces in the counterterm that do not vanish on-shell are gauge invariant \cite{Kallosh}.
\par
In case that the cosmological constant vanishes the final result is
\bea
\Delta S&=&\frac{1}{\e}\,\frac{1}{ (4\pi)^2 } ~\frac{1}{360} ~\int\,d^4 x ~\sqrt{\bg} ~\left\{ 426 \br_{\m\n\r\s}\br^{\m\n\r\s}-1446\br_{\m\n}\br^{\m\n}+435\br^2\right.\nonumber\\
&&\left.+60\br\,\bg^{\r\s}\pd_\r\bp\pd_\s\bp+360\left(\bg^{\r\s}\pd_\r\bp\pd_\s\bp\right)^2+360\bn^2\bp\bn^2\bp\right\}\nonumber\\
&&=\frac{1}{\e}\frac{1}{ (4\pi)^2 } ~\int\,d^n x ~\sqrt{\bg} ~\left\{\frac{43}{60}\br_{\m\n}\br^{\m\n}+\frac{1}{40}\br^2+\frac{1}{6}\br\,\bg^{\r\s}\pd_\r\bp\pd_\s\bp+\left(\bg^{\r\s}\pd_\r\bp\pd_\s\bp\right)^2\right.\nonumber\\
&&\left.+\bn^2\bp\bn^2\bp\right\}
\eea
which coincides with the result of 't Hooft and Veltman except for the last term. That term is however irrelevant in this case since it vanishes due to the background equations of motion. Using them the counterterm can be written in the form
\be
\Delta S=\frac{1}{\e}\frac{1}{ (4\pi)^2 }~\int\,d^4 x ~\sqrt{\bg} ~ \frac{203}{40}~\br^2
\ee
\item On the other hand, if we had considered pure gravity in the presence of a Cosmological Constant that would correspond in our notation to $F_\l(\bp)=1$ and $\bp=0$. Besides, in order to compare with the results present in the literature, we have to subtract from (\ref{count}) the contribution from scalar loops, which is trivially
\bea
a_4^{\phi}&=&\frac{1}{ (4\pi)^{\frac{n}{2}} } ~\frac{1}{360} ~\int\,d^n x ~\sqrt{\bg} ~\left\{2\br_{\m\n\r\s}\br^{\m\n\r\s}-2\br_{\m\n}\br^{\m\n}+5\br^2\right\}\nonumber\\
&&\frac{1}{ (4\pi)^{\frac{n}{2}} } ~\int\,d^n x ~\sqrt{\bg} ~\left\{\frac{1}{180}\br_{\m\n\r\s}\br^{\m\n\r\s}-\frac{1}{180}\br_{\m\n}\br^{\m\n}+\frac{1}{72}\br^2\right\}
\eea
in such a way that, after using the equations of motion and neglecting the topological invariant, the one-loop counterterm coincides with the well known result of Christensen and Duff \cite{Christensen}
\be
\Delta S=\frac{1}{\e}\left(a_4-a_4^\phi-2a_4^{gh}\right)=\frac{1}{\e}\,\frac{1}{ (4\pi)^2 } ~\frac{1}{180} ~\int\,d^n x ~\sqrt{\bg} ~\left\{ 212 \br_{\m\n\r\s}\br^{\m\n\r\s}-2088\,\,\Lambda^2\right\}
\ee
\item It is possible to use the background equations of motion (\ref{beom}) to simplify the final result (\ref{count}). It is convenient to express the counterterm just in terms of the scalar, since we are interested in inverting the conformal transformation. The counterterm is then
\bea
\Delta S&=&\frac{1}{\e}\,\frac{1}{ (4\pi)^2 }~\int\,d^4 x ~\sqrt{\bg} ~\left\{\frac{203}{160}\left(\bg^{\r\s}\pd_\r\bp\pd_\s\bp\right)^2+\frac{57}{20}\,\Lambda\,F_\l(\bp\,)\bg^{\r\s}\pd_\r\bp\pd_\s\bp\right.\nonumber\\
&&\left.-\frac{57}{5}\,\Lambda^2\,(F_\l(\bp))^2+\frac{1}{3}\,\Lambda^2\,(F'_\l(\bp))^2+2\,\Lambda^2\,(F''_\l(\bp))^2-\frac{4}{3}\,\Lambda^2\,F_\l(\bp)\,F''_\l(\bp)\right\}\nonumber\\
\eea
\item If we want to write the counterterm in the original frame we must undo the conformal tranformation, which is very easy once we have it in terms of the scalar. The scalar is related to the conformal factor through (\ref{defsca}), and the conformal factor and the original function of the determinant of the metric verify (\ref{conftrans}). With this in mind, we can express the different contributions to the counterterm in terms of the functions appearing in (\ref{kiner}). Taking into account the definition of the potential $F_\l(\bp)$ given in (\ref{potential}) and supposing that $f(\varphi^*)$ is not a constant we get
\bea
F_\l(\varphi^*)&=&f^{\frac{n}{2-n}}(\varphi^*)\,f_\l(\varphi^*)\nonumber\\
F'_\l(\varphi^*)&=&(n-2)\,f^{\frac{n+2}{4-2n}}\left[\frac{n}{2-n}\,f^{-1}\,f_\l+f'^{-1}\,f'_\l\right]\left[2(n-1)(n-2)\,f^{-1}-(n-2)^2\,f'^{-2}\,f_\phi\right]^{-\frac{1}{2}}\nonumber\\
F''_\l(\varphi^*)&=&(n-2)^2\,f^{\frac{2}{2-n}}\left[2(n-1)(n-2)\,f^{-1}+(n-2)^2\,f'^{-2}\,f_\phi\right]^{-2}\left[\frac{2n^2(n-1)}{n-2}\,f^{-3}\,f_\l\right.\nonumber\\
&&\left.-2(n-1)(n+2)\,f^{-2}\,f'^{-1}\,f'_\l-2(n-1)(n-2)\,f^{-1}\,f'^{-3}\,f''\,f'_\l\right.\nonumber\\
&&\left.+2(n-1)(n-2)\,f^{-1}\,f'^{-2}\,f''_\l-\frac{n(3n-2)}{2}\,f^{-2}\,f'^{-2}\,f_\l\,f_\phi\right.\nonumber\\
&&\left.+\frac{(3n+2)(n-2)}{2}\,f^{-1}\,f'^{-3}\,f'_\l\,f_\phi-\frac{n(n-2)}{2}\,f^{-1}\,f'^{-3}\,f_\l\,f'_\phi\right.\nonumber\\
&&\left.+n(n-2)\,f^{-1}\,f'^{-4}\,f''\,f_\l\,f_\phi-(n-2)^2\,f'^{-4}\,f''_\l\,f_\phi+\frac{(n-2)^2}{2}\,f'^{-4}\,f'_\l\,f'_\phi\right]\nonumber\\
\bg^{\m\n}\pd_\m\bp\pd_\n\bp&=&f^{\frac{2}{2-n}}\left[\frac{2(n-1)}{n-2}\,f^{-2}\,f'^2-f^{-1}\,f_\phi\right]g_*^{\m\n}\pd_\m\varphi^*\pd_\n\varphi^*
\eea

\end{itemize}
               

\begin{thebibliography}{99}

\bibitem{Alvarez}
  E.~Alvarez,
  ``Can one tell Einstein's unimodular theory from Einstein's general
  JHEP {\bf 0503} (2005) 002
  [arXiv:hep-th/0501146].
\\
 E.~Alvarez, D.~Blas, J.~Garriga and E.~Verdaguer,
  ``Transverse Fierz-Pauli symmetry,''
  Nucl.\ Phys.\  B {\bf 756} (2006) 148
  [arXiv:hep-th/0606019].

 
\bibitem{Alvarezfaedo}
  E.~Alvarez and A.~F.~Faedo,
  ``Unimodular cosmology and the weight of energy,''
  Phys.\ Rev.\  D {\bf 76} (2007) 064013
  [arXiv:hep-th/0702184].
  

\bibitem{Comment}
  E.~Alvarez and A.~F.~Faedo,
  ``A comment on the matter-graviton coupling,''
  Phys.\ Rev.\  D {\bf 76} (2007) 124016
  [arXiv:0707.4221 [hep-th]].
  
   
\bibitem{Barvinsky}
  A.~O.~Barvinsky, A.~Y.~Kamenshchik and I.~P.~Karmazin,
  ``The Renormalization Group For Nonrenormalizable Theories: Einstein Gravity With A Scalar Field,''
  Phys.\ Rev.\  D {\bf 48} (1993) 3677
  [arXiv:gr-qc/9302007].
  
  
\bibitem{Barvinskyvilkovisky}
  A.~O.~Barvinsky and G.~A.~Vilkovisky,
  ``The Generalized Schwinger-Dewitt Technique In Gauge Theories And Quantum Gravity,''
  Phys.\ Rept.\  {\bf 119}, 1 (1985).
  
   
\bibitem{Buchmuller}
  W.~Buchmuller and N.~Dragon,
  ``Einstein Gravity From Restricted Coordinate Invariance,''
  Phys.\ Lett.\  B {\bf 207} (1988) 292.
  
  
\bibitem{Christensen}
  S.~M.~Christensen and M.~J.~Duff,
  ``Quantizing Gravity With A Cosmological Constant,''
  Nucl.\ Phys.\  B {\bf 170} (1980) 480.
  
  
\bibitem{DeWitt}
  B.~S.~DeWitt,
  ``Dynamical theory of groups and fields,''
{\it  Gordon \& Breach, New York, 1965}\\
  B.~S.~DeWitt and G.~Esposito,
  ``An introduction to quantum gravity,''
  Int.\ J.\ Geom.\ Meth.\ Mod.\ Phys.\  {\bf 5} (2008) 101
  [arXiv:0711.2445 [hep-th]].
  
  
\bibitem{Dragon}
  N.~Dragon and M.~Kreuzer,
  ``Quantization Of Restricted Gravity,''
  Z.\ Phys.\ C {\bf 41} (1988) 485.
   \\
  M.~Kreuzer,
  ``Gauge Theory Of Volume Preserving Diffeomorphisms,''
  Class.\ Quant.\ Grav.\  {\bf 7} (1990) 1303.

\bibitem{Hooft}
  G.~'t Hooft and M.~J.~G.~Veltman,
  ``One loop divergencies in the theory of gravitation,''
  Annales Poincare Phys.\ Theor.\  A {\bf 20} (1974) 69.


 
\bibitem{Kallosh}
  R.~E.~Kallosh, O.~V.~Tarasov and I.~V.~Tyutin,
  ``One Loop Finiteness Of Quantum Gravity Off Mass Shell,''
  Nucl.\ Phys.\  B {\bf 137} (1978) 145.
  
    

\bibitem{Martin}
xTensor, A fast manipulator of tensor expressions,\\ 
J.~M.~Mart\'in-Garc\'ia 2002-2008,\\ 
(http://metric.iem.csic.es/Martin-Garcia/xAct/) 

  
\bibitem{Skvortsov}
  E.~D.~Skvortsov and M.~A.~Vasiliev,
  ``Transverse invariant higher spin fields,''
  Phys.\ Lett.\  B {\bf 664} (2008) 301
  [arXiv:hep-th/0701278].
  \\
   D.~Blas,
  ``Transverse Symmetry and Spin-3/2 Fields,''
  arXiv:0803.4497 [hep-th].
  
\end{thebibliography}
\end{document}